\documentclass[11pt,twoside]{article}
\usepackage{amsmath,amsthm,amssymb,amsbsy,epsfig,fancyhdr,calc,ifthen,float,psfrag,mathrsfs,hhline,subfigure}
\PassOptionsToPackage{ctagsplt,Righttag}{amstex}
\usepackage[dvips,dvipsnames,usenames]{color} 
\usepackage[colorlinks=true,citecolor=blue,linkcolor=blue,
            bookmarksnumbered=true,pdfstartview={FitH},
            hyperindex=true,linktoc=all,linktocpage=true]{hyperref}
\usepackage{mathtools}
\usepackage{framed}
\usepackage{dsfont}
\usepackage{ulem}
\usepackage{tabularx}
\usepackage{booktabs}
\usepackage[titletoc,title]{appendix}
\usepackage{color}
\definecolor{darkgreen}{rgb}{0,0.50,0}
         
\newcommand{\bb}{\mathbf{b}}
\newcommand{\cb}{\mathbf{c}}

\newcommand{\gb}{\mathbf{g}}

\newcommand{\ib}{\mathbf{i}}

\newcommand{\mb}{\mathbf{m}}

\newcommand{\qb}{\mathbf{q}}

\newcommand{\tb}{\mathbf{t}}

\newcommand{\vb}{\mathbf{v}}
\newcommand{\wb}{\mathbf{w}}
\newcommand{\xb}{\mathbf{x}}
\newcommand{\yb}{\mathbf{y}}

\newcommand{\Ab}{\mathbf{A}}
\newcommand{\Bb}{\mathbf{B}}

\newcommand{\Ib}{\mathbf{I}}
\newcommand{\Jb}{\mathbf{J}}

\newcommand{\Lb}{\mathbf{L}}
\newcommand{\Mb}{\mathbf{M}}

\newcommand{\Qb}{\mathbf{Q}}

\newcommand{\Tb}{\mathbf{T}}

\newcommand{\Bs}{{\mathscr B}}

\newcommand{\Rc}{{\mathcal R}}

\newcommand{\zerob}{\boldsymbol{0}}

\newcommand{\dm}{\mathrm{d}}
\newcommand{\Omegab}{\mathbf{\Omega}}
\newcommand{\omegab}{\boldsymbol{\omega}}

\newcommand{\cP}{{\mathcal P}}
\newcommand{\cR}{{\mathcal R}}

\newcommand{\grady}{\dfrac{\partial}{\partial \yb}}

\newcommand{\gradyp}{\dfrac{\partial}{\partial \yb^+}}
\newcommand{\gradxpg}{\dfrac{\partial g}{\partial \xb^+}}
\newcommand{\gradxg}{\dfrac{\partial g}{\partial \xb}}
\newcommand{\gradypg}{\dfrac{\partial g}{\partial \yb^+}}
\newcommand{\gradyg}{\dfrac{\partial g}{\partial \yb}}

\newcounter{hours}
\newcounter{minutes}
\newcommand{\printtime}{\setcounter{hours}{\time/60}%
                        \setcounter{minutes}{\time-\value{hours}*60}%
\ifthenelse{\value{hours}<10}{0}{}\thehours:%
\ifthenelse{\value{minutes}<10}{0}{}\theminutes}
\def\lsp{\def\baselinestretch{0.75}\large\normalsize}
\def\ssp{\def\baselinestretch{1.0}\large\normalsize}
\def\dsp{\def\baselinestretch{1.37}\large\normalsize} 
\newcommand{\figref}[1]{Figure~\ref{#1}}
\allowdisplaybreaks
\begin{document}
\bibliographystyle{unsrt}
\flushbottom
\pagestyle{empty}
\pagenumbering{arabic}
\ssp 
\title{\vspace{-0.75in}\bf On the Polar Nature and Invariance
Properties \\ of a Thermomechanical Theory \\ for
Continuum-on-Continuum Homogenization \rm}
\vspace{0.1in}
\author{Kranthi K. MANDADAPU$^{1,2}$\footnote{kranthi@berkeley.edu (corresponding 
author)},
B. Emek ABALI$^3$, 
and Panayiotis PAPADOPOULOS$^3$\footnote{panos@berkeley.edu (corresponding 
author)} 
\\[0.05in]
\sl\small $^1$Department of Chemical and Biomolecular Engineering,
University of California, Berkeley, CA 94720-1462, USA \\
{\sl\small $^2$Chemical Sciences Division, Lawrence Berkeley National
Laboratory, Berkeley, CA 94720-1740, USA }\\
\sl\small $^3$Department of Mechanical Engineering, University of
California, Berkeley, CA 94720-1740, USA 
\rm \normalsize}
\date{}
\maketitle
\small
\lsp
\renewcommand{\contentsname}{\normalsize\centerline{Table of contents}}
\tableofcontents
\ssp
\protect\vspace{0.3in}
\normalsize
\begin{abstract}
This paper makes a rigorous case for considering the homogenized continuum
derived by the Irving--Kirkwood procedure as a polar medium in which 
the balances of angular momentum and energy contain contributions due
to body couples and couple stresses defined in terms of the underlying 
microscopic state. The paper also addresses the question of invariance
of macroscopic stress and heat flux and form-invariance of the 
macroscopic balance laws. 
\end{abstract}
\protect\vspace{0.3in}
\nopagebreak[4]
\begin{description}
\item[Keywords:] continuum homogenization; extensivity; Irving-Kirkwood 
procedure; polar media; invariance
\end{description}
\dsp
\newpage
\pagestyle{fancyplain}
%
\section{Introduction}\label{sec:intro}
\par
Continuum-on-continuum homogenization provides a convenient theoretical 
framework for analyzing media in which there exists sufficient
length- and time-separation between the macroscopic body and its
microstructural components, while, at the same time, both may be
accurately modeled as continuous media. This may well be the case for
bulk metals (with their polycrystalline microstructure) and composites
(with, say, their matrix-fiber microstructure). In the general
thermomechanical setting, the goal of homogenization theories is to 
deduce (homogenized) macroscopic counterparts for all the kinematic and kinetic
variables that enter the microscopic description of the continuous medium. 
\par
The pioneering work of Irving and Kirkwood~\cite{Irving1950} on the
upscaling of classical statistical mechanics to 
continuum hydrodynamics motivated a recent study of
continuum-on-continuum homogenization, which
led to the rigorous derivation of formulae for macroscopic stress and heat flux 
based on a minimal set of assumptions, that is, extensivity of mass,
momentum, and energy~\cite{Mand12}. While phase-space averaging
was substituted by mass-weighted volume averaging and interacting
particles in the microscale were replaced by a continuum, 
the critical dependence on extensivity and the procedural similarity in 
the derivations render the continuum-on-continuum homogenization method 
in~\cite{Mand12} a close relative to the original Irving-Kirkwood
method. The resulting formulae incorporate naturally the
volumetric effect of inertia on both stress and heat flux and can
be used in practical computations using, {\em e.g.,} two-scale finite
element methods~\cite{Merc15}.
While it can be plausibly assumed that at appropriately small length
scales such volume effects become negligible compared to surface effects, 
as is argued in the continuum homogenization literature 
(see, {\em e.g.},~\cite{Temi12}), volumetric effects become dominant 
in the presence of non-trivial velocity fluctuations, as is
the case with wave propagation in heterogeneous media where wavelengths 
are in the order of the length scale~\cite{Shen00,Nema11,Merc16}. 
It is important to note that continuum homogenization theories 
based on extensivity have been already considered in other field
theories, such as electrodynamics~\cite{Russakoff70}.
\par
The present paper explores the polar nature of the macroscopic
continuum in the homogenization theory, motivated intuitively by the premise 
that the length scale of the underlying microstructure is generally bound 
to yield non-vanishing body and surface couples. The polar nature is
established methodologically by the approach adopted 
in~\cite{Mandal2017_1,Mandal2017_2} for upscaling atomistic systems 
with internal couples to the continuum hydrodynamics.
In particular, it is shown that the distinction between macroscopic 
angular momentum and moment of momentum, argued masterfully 
in~\cite{Dahler61}, albeit with only a general allusion to directed media, 
is a natural implication of the homogenization theory. In fact, 
it is rigorously confirmed that couple forces, defined in terms
of the microscopic state, enter in a non-trivial statement 
of macroscopic angular momentum balance. The proposed theory
differs from the micromorphic theory~\cite{Eringen1964a,Eringen68} 
both methodologically and philosophically. Indeed, the micromorphic
theory relies on homogenization rules for kinetic quantities,
such as stress and heat flux, which are not extensive. In addition,
constitutive laws for the micromorphic continuum are postulated in the
macroscale without explicit reference to the material constitution or
to geometric features of the underlying microstructure. In contrast, 
the proposed theory relies strictly on homogenization of extensive
quantities and derives the macroscopic constitutive response explicitly 
from the microstructure.
A key further novelty of the proposed analysis is in the kinematics of the
macroscale, which is naturally enriched by an angular velocity quantifying 
the local rotatory effect of the motion and enables the decomposition 
of the kinetic energy into translational and rotational components. 
The angular velocity is related to a macroscopic quantity akin 
to a local moment of inertia, whose evolution is governed 
by its own balance equation. The concept of local moment of inertia 
in a polar medium was considered initially in \cite{eringen1964}, 
where a balance equation is proposed without, however, an associated 
moment of inertia flux term. Other theories of polar media either neglect
the moment of inertia or assume it to be independent of 
time~\cite{frank1958,mindlin1964,leslie1968}.
In contrast, the present theory provides an explicit definition of
a local macroscopic moment of inertia in terms of the microscopic state 
and a corresponding balance law for its evolution that contains a 
moment-of-inertia flux term. 
\par
The paper also addresses the question of invariance in the macroscale
based, again, on a minimal set of assumptions on the form-invariance
of the extensive relations and the underlying microscopic balances. 
Form-invariance of the macroscale balance laws is shown to hold
without any extraneous limitations on the nature of the superposed
rigid motion. Also, the inertial effects on stress and heat flux 
are shown to play a crucial role in the transformation 
of these quantities under superposed rigid-body motions and the 
associated form-invariance of the balance of linear momentum and energy. 
In addition, they may point to a path toward the formal resolution 
of related long-standing controversies on the invariance of stress in 
turbulence~\cite{Luml70,Luml83,Sadi96} and heat flux in rotating 
particle flows~\cite{Muel72,True76,Hoov81}. 
\par
The organization of the paper is as follows: Section~\ref{sec:theory}
contains an outline of the continuum Irving--Kirkwood procedure, as
well as expanded discussion on angular momentum. The homogenization of
total internal energy and its various constituent parts is addressed
in~\ref{sec:energy}, while the matter invariance is investigated in
Section~\ref{sec:FI-extensive} for the principal extensive quantities,
Section~\ref{sec:stress-inv} for stress and linear momentum balance,
Section~\ref{sec:am-inv} for angular momentum, and
Section~\ref{sec:heat-inv} for heat flux and energy balance.
Concluding remarks are offered in Section~\ref{sec:conclusions}.  
\section{Overview of the Extensive Homogenization Method}\label{sec:theory}
\subsection{Review of previous results: balance of mass and linear momentum} 
\par
Consider a body~$\Bs$, which occupies a region~$\Rc$ with boundary
$\partial\Rc$ in the current configuration, and let the positions of material 
points in the microscale and macroscale be denoted~$\xb$ and~$\yb$, 
respectively. Assuming that continuum mechanics is applicable at both
length/time scales, the local forms of the balance laws for mass and linear 
momentum at the microscale may be expressed as 
\begin{equation}\label{eq:micro_mass}
\dot{\rho}^m + \rho^m \dfrac{\partial}{\partial \xb} \cdot \vb^m\ =\ 0\ ,
\end{equation}
\begin{equation}\label{eq:micro_momen}
\rho^m \dot{\vb}^m\ =\ \dfrac{\partial}{\partial \xb} \cdot \Tb^m + 
\rho^m \bb^m\ .
\end{equation}
Likewise, the corresponding balance laws for the macroscale take the form 
\begin{equation}\label{eq:macro_mass}
\dot{\rho}^M + \rho^M \dfrac{\partial}{\partial \yb} \cdot \vb^M \ = \ 0 \ , 
\end{equation}
\begin{equation}\label{eq:macro_momen}
\rho^M \dot{\vb}^M \ = \ \dfrac{\partial}{\partial \yb} \cdot \Tb^M + \rho^M \bb^M \ .
\end{equation} 
Here,~$\rho$ is the mass density,~$\vb$ is the velocity,
$\Tb$ is the Cauchy stress tensor, 
and~$\bb$ is the body force per unit mass. 
In addition, 
``$\frac{\partial}{\partial\yb}\,\cdot$'' and
``$\frac{\partial}{\partial\xb}\,\cdot$'' denote the divergence
operators relative to~$\yb$ and~$\xb$, respectively, while
the overdot denotes material time derivative.
All terms in~(\ref{eq:micro_mass}-\ref{eq:macro_momen}) carry 
a superscript ``$m$'' (for microscale) or ``$M$'' (for macroscale). 
Moreover, microscopic terms are functions of~$(\xb,t)$, while macroscopic 
terms are functions of~$(\yb,t)$. For brevity, explicit declaration 
of these functional dependencies is selectively omitted henceforth. 
\par
Expressions for the macroscopic Cauchy stress and body force are derived by 
postulating homogenization relations for the extensive quantities of mass 
and linear momentum. These are given by 
\begin{equation}\label{eq:ext1}
\rho^M(\yb,t) \ = \ \int_{\cR} \rho^m(\xb,t) g(\yb,\xb) \, \dm v^m \ ,
\end{equation}
\begin{equation}\label{eq:ext2}
\rho^M(\yb,t) \vb^M(\yb,t) \ = \ \int_{\cR} \rho^m(\xb,t) \vb^m(\xb,t) 
g(\yb,\xb)\, \dm v^m \ ,  
\end{equation}
respectively, where~$g(\yb,\xb)$ is a real-valued coarse-graining 
function~\cite{Mand12}. This function is assumed to satisfy the condition 
\begin{equation}\label{eq:g-boundary} 
g(\yb,\xb)\ =\ 0\quad\text{when }\xb\in\partial\Rc
\end{equation}
and also be invariant under superposed rigid-body motions, 
which implies that 
\begin{equation}\label{eq:g} 
g(\yb,\xb)\ =\ g(\yb^+,\xb^+)\ .
\end{equation}
The latter has been shown in~\cite{Mand12} to further imply that 
$g(\yb,\xb) = \bar{g}(|\xb - \yb|)$, hence 
\begin{equation}\label{eq:iden}
\frac{\partial}{\partial \xb}g(\yb,\xb)\ =\ 
-\frac{\partial}{\partial \yb}g(\yb,\xb)\ .
\end{equation}
The support of the coarse-graining function quantifies the length
scale which characterizes the homogenization and is informed by the
geometry and material constitution of the microstructure.
\par
Taking material time derivatives of relations~(\ref{eq:ext1},\ref{eq:ext2}), 
using the balance laws~(\ref{eq:micro_mass},\ref{eq:micro_momen}) 
at the microscale, and comparing the resulting equations to the balance laws 
(\ref{eq:macro_mass},\ref{eq:macro_momen}) at the macroscale, 
the macroscopic Cauchy stress tensor is found in \cite{Mand12} to be 
\begin{equation}\label{eq:macrostress}
\Tb^M\ =\ \int_\cR \bigl[\Tb^m -  \rho^m (\vb^m-\vb^M) \otimes
(\vb^m - \vb^M)\bigr] g\, \dm v^m\ , 
\end{equation}
%
%
%
to within a divergence-free term, while 
%
%
the macroscopic body force is given by 
\begin{equation}\label{eq:bodyM}
\displaystyle \rho^M \bb^M \ = \
\int_{\cR} \rho^m\bb^m g\, \dm v^m \ .
\end{equation}
%
%
%
Equation~\eqref{eq:macrostress} implies that the macroscopic
Cauchy stress is symmetric, as is (on satisfying microscopic angular momentum 
balance) the corresponding microscopic stress. It also demonstrates the
explicit presence of kinetic effects in addition to the (weighted)
average of the microscopic stress. 
\subsection{Homogenization of angular momentum}\label{sec:am} 
\par
The balance of angular momentum at the macroscale is not considered 
in the theory originally proposed in~\cite{Mand12}. In this section, 
the consequences of the homogenization of angular momentum 
are investigated. In particular, the balance of macroscopic angular momentum 
is derived from its microscopic counterpart, and the associated couple stress 
tensor is identified along with the body couple in terms of microscopic 
variables. This process demonstrates the polar nature of the continuum 
homogenization theory proposed in \cite{Mand12}.
\par
Since angular momentum is also an extensive quantity, 
an additional assumption in the continuous Irving--Kirkwood homogenization 
theory is that the total macroscopic angular momentum~$\Lb^M$ per unit mass 
is defined as 
\begin{equation}\label{eq:ext3}
\rho^M(\yb,t) \Lb^M(\yb,t)\ =\ \int_\cR \xb\times \rho^m(\xb,t)
\vb^m(\xb,t) g(\yb,\xb)\, \dm v^m\ . 
\end{equation} 
This can be alternatively expressed with the aid of~\eqref{eq:ext2} as 
\begin{equation}\label{eq:am1}
\rho^M \Lb^M\ =\ \yb\times\rho^M \vb^M + \rho^M \Lb_s^M\ , 
\end{equation} 
where 
\begin{equation}\label{eq:s-am1}
\rho^M \Lb_s^M(\yb,t)\ =\ \int_\cR (\xb-\yb)\times \rho^m(\xb,t)
\vb^m(\xb,t) g(\yb,\xb)\, \dm v^m\ . 
\end{equation}
It is readily concluded from~\eqref{eq:am1} that the total macroscopic 
angular momentum is equal to the macroscopic moment 
of momentum~$\rho^M \yb\times\vb^M$ plus the term~$\rho^M \Lb_s^M$ in
\eqref{eq:s-am1}, which is due to the internal spin in the macroscale, 
see also~\cite{Dahler61}. 
\par
The integral form of angular momentum balance in the macroscale may
now be expressed as 
\begin{equation}\label{eq:amb}
\frac{\dm}{\dm t} \int_\cP \rho^M \Lb^M\,\dm v^M\ =\ 
\int_\cP \yb\times\rho^M\bb^M\,\dm v^M + 
\int_{\partial\cP} \yb\times\tb^M\,\dm a^M +  
\int_\cP \rho^M\gb^M\,\dm v^M + \int_{\partial\cP} \mb^M\,\dm a^M\ ,
\end{equation}
where~$\tb^M$ and~$\mb^M$ denote the macroscopic force and
force couple on the boundary~$\partial\cP$ of an arbitrary region~$\cP$, 
respectively, while~$\gb^M$ is the body couple in~$\cP$. 
Substituting~\eqref{eq:am1} into~\eqref{eq:amb}, applying the Reynolds
transport and divergence theorems, and 
invoking~(\ref{eq:macro_mass},\ref{eq:macro_momen}) and the
symmetry of the Cauchy stress in~\eqref{eq:macrostress}, the statement 
of macroscopic angular momentum balance reduces to 
\begin{equation}\label{eq:s-amb1}
\frac{\dm}{\dm t}\int_\cP \rho^M\Lb_s^M\, \dm v^M\ =\ 
\int_\cP \rho^M\gb^M\,\dm v^M + \int_{\cP}
\frac{\partial}{\partial \yb}\cdot{\Mb^M} \,\dm v^M\ ,
\end{equation}
where~$\Mb^M$ is the couple stress related to the force couple~$\mb^M$
by the standard Cauchy stress theorem. Equation~\eqref{eq:s-amb1} may be
thought of as expressing the balance of the (homogenized) internal spin in the
macroscale. A local macroscopic counterpart of~\eqref{eq:s-amb1}, derived 
directly from the Reynolds transport theorem and~\eqref{eq:macro_mass}, 
takes the form
\begin{equation}\label{eq:s-amb1-local}
\rho^M\dot{\Lb}_s^M\ =\ \rho^M\gb^M + \frac{\partial}{\partial
\yb}\cdot{\Mb^M}\ . 
\end{equation}
\par
Expanding the left-hand side of~\eqref{eq:s-amb1} by employing the
Reynolds transport theorem and taking advantage of~\eqref{eq:micro_mass}, 
\eqref{eq:micro_momen},~(\ref{eq:ext1}-\ref{eq:g-boundary}),~\eqref{eq:iden}, 
the definition in~\eqref{eq:s-am1}, and the symmetry of the microscopic Cauchy 
stress gives rise to    
\begin{multline}\label{eq:amb1b}
\frac{\dm}{\dm t}\int_\cP \rho^M\Lb_s^M\, 
\dm v^M\ =\ 
\int_\cP \left[
\int_\cR (\xb-\yb)\times\rho^m\bb^m g\,\dm v^m \right. \\ 
+
\frac{\partial}{\partial \yb}\cdot\int_\cR 
(\xb-\yb)\times\Tb^m g\,\dm v^m 
\left. 
- \frac{\partial}{\partial \yb}\cdot
\int_\cR (\xb-\yb)\times
\left[\rho^m\vb^m \otimes (\vb^m-\vb^M)\right] g\,\dm v^m
\right]\,\dm v^M\ , 
\end{multline} 
where the (left) cross product between a vector and a second-order
tensor (see, {\em e.g.}, \cite[Section 2.1.7]{Bertram15}) is employed 
in the last two terms of~\eqref{eq:amb1b}. 
Reconciling the right-hand sides of~\eqref{eq:s-amb1} and~\eqref{eq:amb1b}, 
it follows that the macroscopic body couple takes the form 
\begin{equation}\label{eq:couple-f} 
\rho^M\gb^M\ =\ 
\int_\cR (\xb-\yb)\times\rho^m\bb^m g\,\dm v^m\ , 
\end{equation} 
while, to within a divergence-free term, the macroscopic couple stress 
is given by
\begin{equation}\label{eq:couple-m} 
\Mb^M\ =\ \int_\cR (\xb-\yb)\times\Tb^m g\,\dm v^m - 
\int_\cR (\xb-\yb)\times
\left[\rho^m\vb^m \otimes (\vb^m-\vb^M)\right] g\,\dm v^m\ .
\end{equation} 
The term on the right-hand side of~\eqref{eq:couple-f}
is due to the internal torque induced by the microscopic body force.
Likewise, the two terms comprising the (unsymmetric) macroscopic 
couple stress in~\eqref{eq:couple-m} signify the moment of the microscopic 
stress and the fluctuation in the internal spin, respectively. 
\par
It is important to emphasize at this point that the macroscopic angular momentum
balance equations do not represent new physics, but rather
underline the polar nature of the homogenized macroscopic medium
derived from a conventional microscopic continuum. Also, the
macroscopic linear momentum balance equations~\eqref{eq:macro_momen} and 
angular momentum balance equations~\eqref{eq:s-amb1-local} are coupled by
virtue of their dependence on the kinematics and stresses of the
(shared) microstructure.
\section{Homogenization of Energy}\label{sec:energy} 
\par
In view of the polar nature of the macroscopic continuum, 
the homogenization of energy in~\cite{Mand12} 
is reconsidered and alternative expressions are derived for 
the heat supply and heat flux 
by identifying the appropriate forms of the work by couple stress and 
body couple. In addition, an additive decomposition of the total
internal energy is deduced by a suitable definition of the angular
velocity of the macroscopic continuum.
\par
The local form of energy balance in the microscale may be expressed
conventionally as 
\begin{equation}\label{eq:micro_ener}
\rho^m \dot{e}^m\ =\  \rho^m \bb^m \cdot \vb^m + 
\rho^m r^m + \dfrac{\partial}{\partial \xb} \cdot
(\Tb^m  \vb^m )- 
\dfrac{\partial }{\partial \xb} \cdot \qb^m\ ,
\end{equation}
where (upon omitting explicit reference to the 
superscript~``$m$'')~$e = \epsilon + \frac{1}{2} \vb \cdot \vb$ 
is the total internal energy (including kinetic energy) per unit mass, 
with~$\epsilon$ being the internal energy per unit mass,~$\qb$ is the heat-flux 
vector, and~$r$ is the heat supply per unit mass. Also, the symmetry
of the microscopic Cauchy stress has been invoked in deriving the
third term on the right-hand side of~\eqref{eq:micro_ener}. 
The standard reduced form of energy balance in the microscale can be stated as 
\begin{equation}\label{eq:micro_ener-r}
\rho^m \dot{\epsilon}^m\ =\ \rho^m r^m + 
\Tb^m\cdot\frac{\partial\vb^m}{\partial\xb} - 
\dfrac{\partial }{\partial \xb} \cdot \qb^m\ .
\end{equation}
\par
It is tempting to put forth an expression for the macroscopic energy
balance corresponding to~\eqref{eq:micro_ener}, as done previously 
in~\cite{Mand12}. Instead, appreciating the polar nature of the
macroscopic continuum, as demonstrated in
Section~\ref{sec:am}, it is instructive to start 
from the statement of extensivity for the total internal energy, in the form 
\begin{equation}\label{eq:ext4}
\rho^M(\yb,t) e^M(\yb,t) \ = \ 
\int_{\cR} \rho^m(\xb,t) e^m(\xb,t) g(\yb,\xb) \, \dm v^m \ ,
\end{equation}
and explore the full range of its implications in relation to macroscopic 
energy balance. To this end, upon invoking~(\ref{eq:ext1},\ref{eq:ext2})
and the preceding decomposition of the total microscopic internal energy,
equation~\eqref{eq:ext4} readily leads to 
\begin{equation}\label{eq:energy1} 
\rho^M e^M \ = \ 
\int_\Rc \rho^m \epsilon^m g\, \dm v^m + 
\int_\Rc \frac{1}{2} \rho^m (\vb^m-\vb^M)\cdot(\vb^m-\vb^M) g\, \dm v^m + 
\frac{1}{2} \rho^M \vb^M\cdot\vb^M\ .
\end{equation}
Equation~\eqref{eq:energy1} shows that the total macroscopic internal energy 
consists of three distinct parts: the homogenized microscopic internal energy; 
the homogenized kinetic energy of the velocity fluctuations in the microscale; 
and, the macroscopic translational kinetic energy. 
\par
\begin{figure}[ht]
  \centering
  \vskip 2.5in
  \psfragscanon
  \psfrag{x}{$\xb$}
  \psfrag{y}{$\yb$}
  \psfrag{x-y}{$\xb-\yb$}
  \psfrag{w}{$\wb^M$}
  \psfrag{vM}{$\vb^M$}
  \psfrag{vm}{$\vb^m$}
  \psfrag{vhm}{$\hat{\vb}^m$}
  \psfrag{wxy}{$\wb^M\times(\xb-\yb)$}
  \includegraphics[width=.6\textwidth]{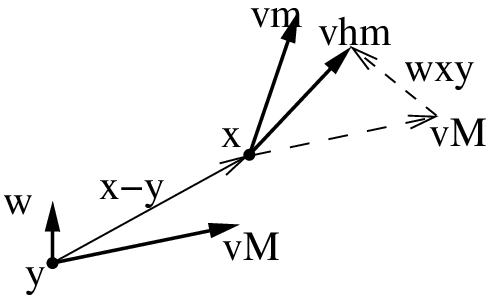}
  \caption{Schematic depiction of the velocities~$\vb^m$,
$\hat{\vb}^m$ and~$\vb^M$ at points with position vectors~$\xb$ and~$\yb$.}
  \label{fig:velo}
\end{figure}

To reveal the central role of spin in the macroscopic energy, let~$\wb^M$ be 
an angular velocity anchored at~$\yb$, and defer its exact prescription until 
later in this section. Next, define the convected microscopic velocity 
$\hat{\vb}^m$ as
\begin{equation}\label{eq:v-convected} 
\hat{\vb}^m\ =\ \vb^M + \wb^M\times(\xb-\yb)\ ,
\end{equation} 
where, in general,~$\hat{\vb}^m\neq\vb^m$, see~\figref{fig:velo}. 
It is now possible to write the kinetic energy of the 
fluctuations in~\eqref{eq:energy1} as  
\begin{multline}\label{eq:kin-fluct} 
\int_\Rc \frac{1}{2} \rho^m (\vb^m-\vb^M)\cdot(\vb^m-\vb^M) g\, \dm v^m\ =\ 
\int_\Rc \frac{1}{2} \rho^m (\vb^m-\hat{\vb}^m)\cdot(\vb^m-\hat{\vb}^m) g\, \dm v^m\\ 
 + \int_\Rc \frac{1}{2}\rho^m
\bigl[\wb^M\times(\xb-\yb)\bigr]\cdot\bigl[\wb^M\times(\xb-\yb)\bigr]g\,\dm v^m
+ \int_\Rc \rho^m
(\vb^m-\hat{\vb}^m)\cdot\bigl[\wb^M\times(\xb-\yb)\bigr] g\, \dm v^m\ . 
\end{multline} 
The second term on the right-hand side of~\eqref{eq:kin-fluct} can be
also expressed as 
\begin{multline}\label{eq:kin-fluct1} 
\int_\Rc \frac{1}{2} \rho^m 
\bigl[\wb^M\times(\xb-\yb)\bigr]\cdot\bigl[\wb^M\times(\xb-\yb)\bigr]g\,
\dm v^m\\ 
=\ \frac{1}{2} \int_\Rc \rho^m 
\bigl[(\xb-\yb)\cdot(\xb-\yb) \ib -
(\xb-\yb)\otimes(\xb-\yb)\bigr]g\,
\dm v^m \cdot (\wb^M\otimes\wb^M)\ =\ \frac{1}{2}\Ib^M\wb^M\cdot\wb^M\ ,
\end{multline} 
in terms of the (homogenized) moment-of-inertia tensor~$\Ib^M$ at point~$\yb$, 
defined classically as
\begin{equation}\label{eq:m-inertia} 
\Ib^M\ =\ \int_\Rc \rho^m 
\bigl[(\xb-\yb)\cdot(\xb-\yb) \ib -
(\xb-\yb)\otimes(\xb-\yb)\bigr]g\, \dm v^m\ ,
\end{equation} 
where~$\ib$ is the spatial second-order identity tensor. 
\par
Starting from~\eqref{eq:m-inertia}, it can be readily
confirmed with the aid of the Reynolds transport 
theorem, as well as equations~\eqref{eq:micro_mass},~\eqref{eq:g-boundary},
and~\eqref{eq:iden} that  
\begin{equation}\label{eq:m-inertia-dot}
\dot{\Ib}^M + \Ib^M \frac{\partial}{\partial \yb}\cdot\vb^M
+ \frac{\partial}{\partial\yb}\cdot\Jb^M\ =\ \zerob\ ,
\end{equation} 
where 
\begin{equation}\label{eq:J} 
\Jb^M\ =\ \int_\Rc\rho^m \bigl[(\xb-\yb)\cdot(\xb-\yb) \ib -
(\xb-\yb)\otimes(\xb-\yb)\bigr] \otimes(\vb^m-\vb^M)g\, \dm v^m
\end{equation}
is a third-order macroscopic moment-of-inertia flux tensor. 
Equation~\eqref{eq:m-inertia-dot} expresses the (derivable rather 
than primitive) balance of the moment of inertia and stands in
qualitative contrast to the conventional macroscopic mass balance 
equation~\eqref{eq:macro_mass}. In particular, it demonstrates that
there is non-material transport of rotational inertia owing to 
the fluctuations in the velocity, as evidenced by the third term 
on the left-hand side of~\eqref{eq:m-inertia-dot}. 
\par
The last term on the right-hand side of~\eqref{eq:kin-fluct} can be
written with the aid of~\eqref{eq:v-convected} and~\eqref{eq:m-inertia} as 
\begin{equation}\label{eq:kin-fluct2} 
\int_\Rc \rho^m
(\vb^m-\hat{\vb}^m)\cdot\bigl[\wb^M\times(\xb-\yb)\bigr] g\, \dm v^m\
=\  
\int_\Rc \rho^m
(\xb-\yb)\times(\vb^m-\vb^M) g\, \dm v^m\cdot\wb^M - \Ib^M\wb^M\cdot\wb^M\ .
\end{equation} 
Note that the first term on the right-hand side of~\eqref{eq:kin-fluct2}
involves the spin angular momentum in~\eqref{eq:s-am1},
only considered relative to the macroscopic velocity~$\vb^M$. It is now
possible to {\em define\/} the angular velocity~$\wb^M$ such that the
left-hand side of~\eqref{eq:kin-fluct2} vanish identically, which would
imply that 
\begin{equation}\label{eq:w} 
\int_\Rc \rho^m (\xb-\yb)\times(\vb^m-\vb^M) g\, \dm v^m\ =\ \Ib^M\wb^M\ .
\end{equation} 
Therefore,~$\wb^M$ may be thought of as the angular velocity at~$\yb$ which,
when pre-multiplied by the (local) moment-of-inertia tensor, quantifies 
the internal spin relative to the macroscopic velocity. 
\par
The preceding definition effectively eliminates the coupling between the
translational and rotational velocity in the internal energy
of~\eqref{eq:energy1}. Indeed, taking into account 
equations~\eqref{eq:kin-fluct},~\eqref{eq:kin-fluct1}, 
and~\eqref{eq:kin-fluct2}, in connection with the definition of~$\wb^M$ 
in~\eqref{eq:w}, the total macroscopic internal energy in~\eqref{eq:energy1} 
now takes the additive form 
\begin{equation}\label{eq:energy2} 
\rho^M e^M \ = \ 
\rho^M \epsilon^M + 
\frac{1}{2} \rho^M \vb^M\cdot\vb^M + 
\frac{1}{2} \Ib^M \wb^M\cdot\wb^M\ ,
\end{equation}
where the macroscopic internal energy~$\epsilon^M$ per unit mass 
is defined as
\begin{equation}\label{eq:m-int-energy} 
\rho^M \epsilon^M\ =\ \int_\Rc \rho^m \epsilon^m g\, \dm v^m + 
\int_\Rc \frac{1}{2} \rho^m
(\vb^m-\hat{\vb}^m)\cdot(\vb^m-\hat{\vb}^m) g\, \dm v^m\ . 
\end{equation}
The last two terms in~\eqref{eq:energy2} correspond respectively 
to the macroscopic kinetic energy due to translational and rotational effects. 
Moreover,~$\rho^M \epsilon^M$ in~\eqref{eq:m-int-energy} is the
macroscopic internal energy due to all sources other than (macroscopic) 
kinetic energy and includes the effect of kinetic energy fluctuations 
relative to the convected velocity~$\hat{\vb}^m$, which are understood 
here as a manifestation of thermal, rather than mechanical, energy.
\par
Starting from the extensivity relation~\eqref{eq:ext4}, a local
statement of macroscopic energy balance may be derived (see
Appendix~\ref{sec:app-A}) in the form
\begin{multline}\label{eq:ma-energy-balance} 
\rho^M \dot{e}^M\ =\  
\rho^M \bb^M \cdot \vb^M + 
\rho^M \gb^M \cdot \wb^M + 
\int_\Rc\bigl[\rho^m r^m + \rho^m\bb^m\cdot(\vb^m-\hat{\vb}^m)\bigr]g\,\dm v^m
+ \\
\dfrac{\partial}{\partial \yb} \cdot (\Tb^M  \vb^M ) + 
\dfrac{\partial}{\partial \yb} \cdot \left[(\Mb^M)^T  \wb^M \right] - 
\dfrac{\partial }{\partial \yb} \cdot 
\int_\Rc \left[\qb^m - \Tb^m(\vb^m-\hat{\vb}^m) + 
\rho^m \hat{e}^m (\vb^m-\vb^M)\right]g\,\dm v^m\ ,
\end{multline}
where
\begin{equation}\label{eq:e-hat}
\hat{e}^m\ =\ \epsilon^m + 
\frac{1}{2}(\vb^m-\hat{\vb}^m)\cdot(\vb^m-\hat{\vb}^m) - 
\frac{1}{2}\hat{\vb}^m\cdot\hat{\vb}^m\ ,
\end{equation} 
and the superscript ``$T$'' signifies tensorial transpose. 
In contrast to its microscopic counterpart in~\eqref{eq:micro_ener},
equation~\eqref{eq:ma-energy-balance} contains power terms 
involving the body couple~$\gb^M$ and the couple stress~$\Mb^M$,
thus further demonstrating the polar nature of the homogenized macroscopic 
continuum. Also, as seen directly from~\eqref{eq:e-hat}, 
the quantity~$\hat{e}^m$ comprises two competing energetic contributions: 
first, the microscopic internal energy including the kinetic energy of the 
fluctuations of the microscopic velocity relative to the 
``rigid-body motion'' induced (locally) by~$\vb^M$ and~$\wb^M$, and
second, the kinetic energy of the same motion. 
\par
The structure of equation~\eqref{eq:ma-energy-balance} implies that 
the macroscopic heat supply may be defined as
\begin{equation}\label{eq:m-heat-supply}
\rho^M r^M\ =\
\int_\Rc\bigl[\rho^m r^m + \rho^m\bb^m\cdot(\vb^m-\hat{\vb}^m)\bigr]g\,\dm v^m\ , 
\end{equation} 
while, to within a divergence-free term, the macroscopic heat flux is
given by 
\begin{equation}\label{eq:m-heat-flux}
\qb^M\ =\
\int_\Rc \left[\qb^m - \Tb^m(\vb^m-\hat{\vb}^m) + 
\rho^m \hat{e}^m (\vb^m-\vb^M)\right]g\,\dm v^m\ .
\end{equation}
With the preceding definitions in place, the local statement of
macroscopic energy balance~\eqref{eq:ma-energy-balance} takes the form
\begin{equation}\label{eq:ma-eb} 
\rho^M \dot{e}^M\ =\  
\rho^M \bb^M \cdot \vb^M + 
\rho^M \gb^M \cdot \wb^M + 
\rho^M r^M + 
\dfrac{\partial}{\partial \yb} \cdot (\Tb^M  \vb^M ) + 
\dfrac{\partial}{\partial \yb} \cdot \left[(\Mb^M)^T  \wb^M \right] - 
\dfrac{\partial }{\partial \yb} \cdot \qb^M\ .
\end{equation}
\par
It is important to observe here that the definitions 
in~(\ref{eq:m-heat-supply},\ref{eq:m-heat-flux}) and the energy
balance statement in~\eqref{eq:ma-eb} readily reduce to those derived 
in~\cite{Mand12} upon neglecting the angular velocity~$\wb^M$, which 
is tantamount to outright suppressing the polar effects in the macroscale.
%
\section{Invariance of Extensive Relations}\label{sec:FI-extensive}
\par
In this section, the question of invariance under superposed rigid-body motions 
is investigated systematically for the relations~\eqref{eq:ext1},
\eqref{eq:ext2},~\eqref{eq:ext3}, and~\eqref{eq:ext4} 
between the principal extensive quantities in the two scales. 
\par
By way of background, recall that, under superposed rigid-body motions, 
any macroscopic point~$\yb$ in the current configuration of~$\Rc$ is 
mapped to 
\begin{equation}\label{eq:srbm}
\yb^+ \ = \ \Qb\yb + \cb \ ,
\end{equation}
where~$\Qb$ is an arbitrary time-dependent rotation tensor 
and~$\cb$ is an arbitrary time-dependent 
translation vector. It follows from~\eqref{eq:srbm} that the
corresponding velocity and acceleration of this point are given by 
\begin{equation}\label{eq:srbm_v}
{\vb^M}^+ \ = \ \Qb\vb^M + \dot{\Qb} \yb + \dot{\cb} \ ,
\end{equation}
\begin{equation}\label{eq:srbm_a}
\mbox{${{\dot{\vb}}^M}$}^+ \ =\ 
\Qb\dot{\vb}^M + 2\dot{\Qb}\vb^M + \ddot{\Qb}\yb + \ddot{\cb} \ ,
\end{equation}
respectively. In complete analogy to~(\ref{eq:srbm}--\ref{eq:srbm_a}),
one may express the position, velocity and acceleration of any
microscopic material point~$\xb$ under the same superposed rigid-body motion as 
\begin{equation}\label{eq:m-srbm}
\xb^+ \ = \ \Qb\xb + \cb \ ,
\end{equation}
\begin{equation}\label{eq:m-srbm_v}
{\vb^m}^+ \ = \ \Qb\vb^m + \dot{\Qb} \xb + \dot{\cb} \ ,
\end{equation}
\begin{equation}\label{eq:m-srbm_a}
\mbox{${{\dot{\vb}}^m}$}^+ \ =\ 
\Qb\dot{\vb}^m + 2\dot{\Qb}\vb^m + \ddot{\Qb}\xb + \ddot{\cb} \ ,
\end{equation}
respectively. 
\par
At this stage, it is postulated that the relations 
\eqref{eq:ext1},~\eqref{eq:ext2},~\eqref{eq:am1}, and~\eqref{eq:ext4} 
which connect the two scales must be form-invariant under superposed rigid-body 
motions (in the sense of \cite{Green95,Svendsen99}).
This reflects the physically plausible idea that extensive quantities 
should remain extensive under superposed rigid-body motion.
Furthermore, it is assumed that the microscopic balance laws 
in~\eqref{eq:micro_mass},~\eqref{eq:micro_momen},~\eqref{eq:micro_ener}
are likewise form-invariant under superposed rigid-body motions.
\par
Starting with~\eqref{eq:ext1}, form-invariance implies that 
\begin{equation}\label{eq:ext1-srbm}
{\rho^M}^+(\yb^+,t) \ = \ \int_{\cR^+} {\rho^m}^+(\xb^+,t) g(\yb^+,\xb^+)\,
\dm {v^m}^+ \ .
\end{equation}
Upon recalling that~${\rho^m}^+ = \rho^m$ is necessary to ensure
form-invariance of the microscopic mass balance, noting that 
volume are unchanged under superposed
rigid-body motions in the microscale (that is,~$\dm {v^m}^+= \dm {v^m}$),
and also using~\eqref{eq:g}, 
it follows immediately from~\eqref{eq:ext1-srbm} that
\begin{equation}\label{eq:mass-inv} 
{\rho^M}^+\ =\ \rho^M\ .
\end{equation}
\par
Form-invariance of the extensive relation~\eqref{eq:ext2} for linear
momentum necessitates that 
\begin{equation}\label{eq:ext2-srbm}
{\rho^M}^+(\yb^+,t)\, {\vb^M}^+(\yb^+,t)\ =\ \int_{\cR^+} {\rho^m}^+ {\vb^m}^+ g(\yb^+,\xb^+)\, 
\dm {v^m}^+ \ .
\end{equation}
Substituting the expressions for the macroscopic and microscopic velocities 
from~\eqref{eq:srbm_v} and~\eqref{eq:m-srbm_v}, appealing to
\eqref{eq:ext2}, and using, again,~\eqref{eq:g} and the
invariance of density and infinitesimal volume in the microscale, 
equation~\eqref{eq:ext2-srbm} yields 
\begin{equation}\label{eq:cenmass0}
\dot{\Qb}\Big(\rho^{M} \yb -
\int_{\cR} \rho^{m}\xb g\,\dm v^{m} \Big) +
\dot{\cb}\Big( \rho^M - \int_{\cR} \rho^{m} g\,\dm v^{m} \Big)
\ = \zerob\ .
\end{equation}
Setting~$\dot{\Qb} = \zerob$ in~\eqref{eq:cenmass0}, it follows from
the arbitrariness of~$\dot{\cb}$ that the homogenization relation 
\eqref{eq:ext1} for mass may be derived (rather than assumed at the outset) 
from the invariance of the homogenization relation~\eqref{eq:ext2} for 
linear momentum. 
Moreover, upon defining the skew-symmetric 
tensor~$\Omegab = \Qb^T\dot{\Qb}$ and its associated axial vector~$\omegab$,
it follows from the reduced form of~\eqref{eq:cenmass0} and the
arbitrariness of~$\Qb$ that 
\begin{equation}\label{eq:cenmass}
\omegab \times  \Big(\rho^{M} \yb  - 
\int_{\cR} \rho^{m}\xb g(\yb,\xb) \, \dm v^{m} \Big) \ = \ \zerob \ .
\end{equation}
Since~\eqref{eq:cenmass} is valid for all vectors~$\omegab$, it is 
concluded that 
\begin{equation}\label{eq:cenmass-1}
\rho^{M} \yb\ =\ \int_{\cR} \rho^{m}\xb g(\yb,\xb) \, \dm v^{m}\ ,
\end{equation}
which necessitates that the macroscopic point~$\yb$ be located at the 
($g$-weighted) center of mass of the microscopic region around~$\yb$. 
Furthermore, starting from~\eqref{eq:cenmass-1}, 
it can be readily shown with the aid of~\eqref{eq:srbm},
\eqref{eq:m-srbm},~\eqref{eq:mass-inv}, together with~\eqref{eq:ext1}, 
\eqref{eq:g}, and the invariance of microscopic density and volume that 
\begin{equation}\label{eq:cenmass-2}
{\rho^{M}}^+ \yb^+\ =\ \int_{\cR^+} {\rho^{m}}^+ {\xb}^+ g(\yb^+,\xb^+) \, 
\dm {v^{m}}^+\ ,
\end{equation}
therefore the center-of-mass condition~\eqref{eq:cenmass-1} is itself
form-invariant. This condition is of practical importance in computations, 
where its violation would lead to compounding errors, a point which is already 
well-recognized in the related molecular dynamics literature~\cite{Chiu00}.  
An immediate implication of~\eqref{eq:cenmass-1} is that 
the spin angular momentum in \eqref{eq:s-am1} now coincides with its 
counterpart relative to the center of mass, which enters the definition 
of the angular velocity~$\wb^M$ in~\eqref{eq:w}.
A further implication of~\eqref{eq:cenmass-1}, in conjunction with
\eqref{eq:J} is in restating the macroscopic balance of angular momentum 
equation~\eqref{eq:s-amb1-local} in terms of the angular velocity~$\wb^M$ as 
\begin{equation}\label{eq:s-amb2-local}
\Ib^M\dot{\wb}^M\ =\ \rho^M\gb^M + \frac{\partial}{\partial \yb}\cdot \Mb^M + 
\left(\frac{\partial}{\partial \yb}\cdot \Jb^M\right)\wb^M\ ,
\end{equation}
with the last term on the right-hand side reflecting, again, the
effect of the non-material transport of rotational inertia. 
Moreover, upon also taking advantage of~\eqref{eq:m-int-energy}, as well 
as of \eqref{eq:macro_mass},~\eqref{eq:macro_momen},~\eqref{eq:m-inertia-dot}, 
and~\eqref{eq:s-amb2-local}, the reduced form of the energy balance equation 
in the macroscale is easily derived from~\eqref{eq:ma-eb} as 
\begin{equation}\label{eq:ma-eb-r} 
\rho^M \dot{\epsilon}^M\ =\  
\rho^M r^M + 
\Tb^M\cdot\dfrac{\partial\vb^M}{\partial \yb} + 
\Mb^M\cdot\dfrac{\partial\wb^M}{\partial \yb} - 
\frac{1}{2}\left(\frac{\partial}{\partial \yb}\cdot \Jb^M\right)\wb^M\cdot\wb^M 
- \dfrac{\partial }{\partial \yb} \cdot \qb^M\ .
\end{equation}
Again, it is instructive to compare~\eqref{eq:ma-eb-r} to its
microscopic counterpart~\eqref{eq:micro_ener-r}. 
\par
Proceeding to angular momentum, form-invariance of the extensive 
relation~\eqref{eq:am1} implies that 
\begin{equation}\label{eq:am+} 
{\rho^M}^+ {\Lb^M}^+\ =\ \yb^+\times{\rho^M}^+ {\vb^M}^+ + {\rho^M}^+ 
{\Lb_s^M}^+\ . 
\end{equation}
This does not yield additional restrictions on any kinematic or kinetic
variables. Rather, it furnishes an explicit relation between the 
angular momenta~$\rho^M \Lb^M$ and~${\rho^M}^+{\Lb^M}^+$. To derive 
this relation, start with the spin angular momentum term in~\eqref{eq:am+} and
observe, using~\eqref{eq:s-am1},~\eqref{eq:m-inertia},~\eqref{eq:srbm}, 
\eqref{eq:m-srbm},~\eqref{eq:m-srbm_v},~\eqref{eq:cenmass-1}, as well as the
invariance of microscopic density and volume, that 
\begin{align}\label{eq:spin-am-inv}
{\rho^M}^+ {\Lb_s^M}^+ & \ =\ 
\int_{\cR^+} ({\xb}^+-{\yb}^+)\times {\rho^m}^+ {\vb^m}^+ g^+\, \dm
{v^m}^+ \notag \\
&\ =\ 
\int_\cR \bigl[\Qb(\xb-\yb)\bigr]\times \rho^m (\Qb\vb^m +
\dot{\Qb}\xb + \dot{\cb}) g\, \dm v^m \notag \\[0.1in]
&\ =\ 
\Qb\rho^M \Lb_s^M + \Qb\Ib^M\omegab\ ,
\end{align} 
where, for brevity,~$g^+ = g(\yb^+,\xb^+)$.
The preceding relation shows that the deviation of spin
angular momentum from invariance equals an (additive) contribution due to
the angular velocity~$\omegab$ of the superposed rigid-body motion. 
Substituting~\eqref{eq:spin-am-inv} into~\eqref{eq:am+} and
recalling~\eqref{eq:srbm},~\eqref{eq:srbm_v} and~\eqref{eq:mass-inv},
it follows that 
\begin{multline}\label{eq:am-inv} 
{\rho^M}^+ {\Lb^M}^+\ =\ \Qb \rho^M \Lb^M + 
\Qb\Bigl[\rho^M\bigl[(\yb\cdot\yb)\ib - \yb\otimes\yb\bigr] + \Ib^M\Bigr]\omegab \\ + 
(\Qb\yb + \cb)\times\rho^M\dot{\cb} + 
\cb\times\rho^M\Qb(\vb^M + \omegab\times\yb)\ .
\end{multline} 
As seen from~\eqref{eq:am-inv}, additional angular momentum is generated 
by the superposed angular velocity~$\omegab$, the superposed translational 
velocity~$\dot{\cb}$ and the coupling of the macroscopic velocity 
with the superposed translation and rotation.
\par
Lastly, imposing form-invariance to the energy relation~\eqref{eq:ext4}, as
further expanded in~\eqref{eq:energy2} and~\eqref{eq:m-int-energy},
leads to 
\begin{equation}\label{eq:energy+}
{\rho^M}^+ {e^M}^+ \ = \ 
{\rho^M}^+ {\epsilon^M}^+ + 
\frac{1}{2} {\rho^M}^+ {\vb^M}^+\cdot{\vb^M}^+ + 
\frac{1}{2} {\Ib^M}^+ {\wb^M}^+\cdot{\wb^M}^+\ ,
\end{equation}
where
\begin{equation}\label{eq:int-energy+} 
{\rho^M}^+ {\epsilon^M}^+\ =\ \int_{\Rc^+} {\rho^m}^+ {\epsilon^m}^+
g^+\, \dm {v^m}^+ + 
\int_{\Rc^+} \frac{1}{2} {\rho^m}^+
({\vb^m}^+ - {\mbox{$\hat{\vb}^m$}}^+)\cdot 
({\vb^m}^+ - {\mbox{$\hat{\vb}^m$}}^+) g^+\, \dm {v^m}^+\ .
\end{equation}
To start exploring the implications of~\eqref{eq:energy+}, observe that
\begin{align}\label{eq:I+} 
{\Ib^M}^+ & \ =\ \int_{\Rc^+} {\rho^m}^+ 
\bigl[({\xb}^+-{\yb}^+)\cdot({\xb}^+-{\yb}^+) \ib -
({\xb}^+-{\yb}^+)\otimes({\xb}^+-{\yb}^+)\bigr]g^+\, \dm {v^m}^+ \notag \\
& \ =\ 
\int_\Rc \rho^m
\Bigl[\bigl[\Qb(\xb-\yb)\bigr]\cdot\bigl[\Qb(\xb-\yb)\bigr] \ib -
\bigl[\Qb(\xb-\yb)\bigr]\otimes\bigl[\Qb(\xb-\yb)\bigr]\Bigr]g\, \dm {v^m} \notag \\
& \ =\ \Qb\Ib^M\Qb^T\ ,
\end{align}
a well-known result in rigid-body dynamics, 
which follows from~\eqref{eq:srbm},~\eqref{eq:m-srbm},~\eqref{eq:g}, and
the invariance of microscopic density and volume. Next, upon taking advantage
of the definition~\eqref{eq:w}, written under superposed rigid-body
motion as 
\begin{equation}\label{eq:Iw+} 
\int_{\Rc^+} {\rho^m}^+ ({\xb}^+-{\yb}^+)\times({\vb^m}^+-{\vb^M}^+)
g^+\, \dm {v^m}^+\ =\ {\Ib^M}^+{\wb^M}^+\ ,
\end{equation}
one may relate the angular velocity~$\wb^M$ to its counterpart~${\wb^M}^+$. 
To wit, 
\begin{align}\label{eq:Iw+1} 
{\Ib^M}^+{\wb^M}^+ & \ =\ 
\int_\Rc \rho^m \bigl[\Qb(\xb-\yb)\bigr]\times\bigl[\Qb(\vb^m-\vb^M) +
\dot{\Qb}(\xb-\yb)\bigr] g\, \dm v^m \notag \\
& \ =\ 
\Qb \int_\Rc \rho^m (\xb-\yb)\times(\vb^m-\vb^M) g\, \dm v^m + 
\Qb \int_\Rc \rho^m (\xb-\yb)\times\Omegab(\xb-\yb) g\, \dm v^m \notag \\[0.1in]
& \ =\ 
\Qb\Ib^M\wb^M + \Qb\Ib^M\omegab\ ,
\end{align} 
upon invoking~\eqref{eq:m-inertia} and, once again,~\eqref{eq:w}.
This, in conjunction with~\eqref{eq:I+}, implies that 
\begin{equation}\label{eq:w+} 
{\wb^M}^+\ =\ \Qb(\wb^M + \omegab)\ ,
\end{equation}
which reveals the additive effect of the superposed angular velocity on~$\wb^M$.
It now follows from~\eqref{eq:srbm}, \eqref{eq:srbm_v},~\eqref{eq:m-srbm}, 
\eqref{eq:m-srbm_v}, and~\eqref{eq:w+} that, under superposed rigid-body 
motions, the relative velocity~$\vb^m-\hat{\vb}^m$ transforms as 
\begin{align}\label{eq:v-vh+}
{\vb^m}^+ - {\mbox{$\hat{\vb}^m$}}^+ & \ =\ 
{\vb^m}^+ - {\vb^M}^+ - {\wb^M}^+\times(\xb^+-\yb^+) \notag \\
& \ =\ 
\Qb(\vb^m-\vb^M) + \dot{\Qb}(\xb-\yb) -
\Qb(\wb^M+\omegab)\times\Qb(\xb-\yb)  \notag \\[0.04in]
& \ =\ \Qb(\vb^m - \hat{\vb}^m)\ , 
\end{align} 
which proves that this term (unlike~$\vb^m-\vb^M$) is objective. 
Next, recalling that the assumed form-invariance of the microscopic 
energy balance~\eqref{eq:micro_ener-r} is satisfied provided that 
\begin{equation}\label{eq:fi-energy}
{\epsilon^m}^+\ =\ {\epsilon^m}\quad,\quad
{r^m}^+\ =\ {r^m}\quad,\quad
{\qb^m}^+\ =\ \Qb\qb^m\ ,
\end{equation}
it can be shown starting from~\eqref{eq:int-energy+}, with the aid of
\eqref{eq:v-vh+} and~\eqref{eq:fi-energy}$_1$, that 
\begin{equation}\label{eq:int-energy-inv} 
{\rho^M}^+ {\epsilon^M}^+\ =\ \rho^M \epsilon^M\ .
\end{equation}
This means that the macroscopic internal energy (including the kinetic energy
of the velocity fluctuations relative to~$\hat{\vb}^m$) is unaffected 
by superposed rigid-body motions, a result which is highly desirable 
on physical grounds. A straightforward calculation shows that the total 
internal energy in~\eqref{eq:energy+} relates to its counterpart
before the superposition of a rigid-body motion according to 
\begin{multline}\label{eq:energy-inv} 
{\rho^M}^+ {e^M}^+\ =\ \rho^M e^M + 
\frac{1}{2} \rho^M (\omegab\times\yb)\cdot(\omegab\times\yb) + 
\frac{1}{2} \rho^M \dot{\cb}\cdot\dot{\cb} + 
\frac{1}{2} \Ib^M \omegab\cdot\omegab \\[0.05in]
+\dot{\cb}\cdot \rho^M\Qb(\vb^M + \omegab\times\yb) 
+\omegab\cdot (\rho^M \yb\times\vb^M + \Ib^M\wb^M)\ ,
\end{multline}
with each of the additional terms on the right-hand side corresponding
to contributions due to the superposed rigid translation and rotation. 
\section{Invariance: Macroscopic Cauchy Stress and Linear Momentum
Balance}\label{sec:stress-inv} 
\par
Under a superposed rigid-body motion, the macroscopic Cauchy stress 
of~\eqref{eq:macrostress} becomes 
\begin{equation}\label{eq:macrostress_srbm}
{\Tb^{M}}^+\ =\ \int_\cR \bigl[{\Tb^{m}}^+ - {\rho^m}^+
({\vb^{m}}^+ -{\vb^{M}}^+) \otimes ({\vb^{m}}^+ - 
{\vb^M}^+)\bigr] g^+\, \dm {v^{m}}^+\ .
\end{equation}
Assuming the usual invariance relation~$\Tb^{m^+} = \Qb \Tb^m
\Qb^T$ at the microscale, see, {\em e.g.,}~\cite{Chadwick}, 
appealing to the invariance of microscopic density and volume, 
and exploiting~\eqref{eq:g} and the transformation equations~\eqref{eq:srbm_v} 
and~\eqref{eq:m-srbm_v} for the velocity,~\eqref{eq:macrostress_srbm} yields
\begin{align}\label{eq:macrostress_srbm_1}
{\Tb^M}^+\ & =\ 
\int_\cR 
\Big[\Qb \Tb^m \Qb^T - 
 \rho^{m} \bigl[\Qb(\vb^m-\vb^M) + \dot{\Qb}(\xb-\yb)\bigr]
\otimes \bigl[\Qb(\vb^m-\vb^M) +\dot{\Qb}(\xb-\yb)
\bigr]\Big]g\, \dm v^{m} \notag \\
& =\ \Qb 
\biggl[\int_\cR 
\bigl[\Tb^m  -  \rho^{m} (\vb^m-\vb^M) \otimes (\vb^m-\vb^M)\bigr]g\, \dm v^{m} 
\biggr]\Qb^T 
-\int_\cR \rho^m \Qb(\vb^m-\vb^M) \otimes
\dot{\Qb}(\xb-\yb)  g\, \dm v^{m} \notag \\
& \hspace{0.25in} -\int_\cR \rho^m \dot{\Qb}(\xb-\yb) \otimes \Qb(\vb^m-\vb^M) 
g\, \dm v^{m} 
- \int_\cR \rho^m \dot{\Qb}(\xb-\yb) \otimes \dot{\Qb}(\xb-\yb)  g\, \dm v^{m}\ . 
\end{align}
Recalling~\eqref{eq:macrostress} and the definition of~$\Omegab$, 
the preceding equation may be rewritten compactly as  
\begin{equation}\label{eq:macrostress_srbm_3}
{\Tb^M}^+\ =\ 
\Qb \Tb^M \Qb^T - \Qb\left[\Omegab\Ab + (\Omegab\Ab)^T + 
\Omegab\Bb\Omegab^T\right]\Qb^T\ ,
\end{equation}
where 
\begin{equation}\label{eq:A}
\Ab\ =\ \int_\cR {\rho^m} (\xb - \yb) \otimes (\vb^m - \vb^M) g\,\dm v^m 
\end{equation}
and
\begin{equation}\label{eq:B}
\Bb\ =\ \int_\cR \rho^m (\xb-\yb)\otimes (\xb-\yb) g\, \dm v^m\ .
\end{equation}
Note that the tensor~$\Bb$ is symmetric, hence the symmetry of the macroscopic
Cauchy stress in~\eqref{eq:macrostress_srbm_3} is preserved. 
\par
It is clear from~\eqref{eq:macrostress_srbm_3} that the transformation
of the macroscopic Cauchy stress under superposed rigid-body motions does 
not generally obey the conventional continuum mechanics invariance relation. 
In fact, all of the additional terms on the right-hand side 
of~\eqref{eq:macrostress_srbm_3} involve the angular velocity~$\Omegab$. 
Indeed, the first (symmetrized) pair of terms reflects the contribution 
of the angular momentum due to the fluctuations~$\vb^m - \vb^M$, while the
last term quantifies the effect of the unit cell's moment of inertia 
on the macroscopic stress. It is shown in Appendix~\ref{sec:app-B} that these 
additional terms are individually 
divergence-free with respect to the macroscopic coordinates of the
system in the rigidly transformed frame, hence they do not affect the
balance of linear momentum in that frame. Clearly, if the
contribution of the velocity fluctuation terms
in~\eqref{eq:macrostress} is negligible (which would be a reasonable
assumption for most problems involving solids), objectivity 
of the Cauchy stress tensor is restored.
\par
An important additional implication of the divergence-free property 
of the non-invariant terms in~\eqref{eq:macrostress_srbm_3} 
is in the question of form-invariance for the macroscopic linear momentum 
balance. Indeed, recalling that, in view of~\eqref{eq:srbm_a}, form-invariance 
of the microscopic linear momentum balance translates to the condition 
\begin{equation}\label{eq:b+}
\rho^m{\bb^m}^+\ =\ \Qb\rho^m\bb^m + 
\rho^m(2 \dot{\Qb}\vb^m + \ddot{\Qb}\xb + \ddot{\cb})\ , 
\end{equation}
one may readily conclude with the aid of~(\ref{eq:macro_momen}--\ref{eq:ext2}), 
\eqref{eq:bodyM},~\eqref{eq:srbm_a},~\eqref{eq:m-srbm_a},~\eqref{eq:mass-inv}, 
\eqref{eq:cenmass-1}, and~\eqref{eq:macrostress_srbm_3} that 
\begin{equation}\label{eq:blm-fi}
{\rho^M}^+ \mbox{${\dot{\vb}^M}$}^+ \ = \ \dfrac{\partial}{\partial
\yb^+} \cdot {\Tb^M}^+ + {\rho^M}^+ {\bb^M}^+\ ,
\end{equation} 
where
\begin{equation}\label{eq:b-ma+}
{\rho^M}^+ {\bb^M}^+\ =\
\int_{\cR^+} {\rho^m}^+{\bb^m}^+ g^+\, \dm {v^m}^+\ =\ 
\Qb\rho^M\bb^M + \rho^M(2 \dot{\Qb}\vb^M + \ddot{\Qb}\yb + \ddot{\cb})\ .
\end{equation}
\par
The importance of the transformation condition~\eqref{eq:macrostress_srbm_3} 
is alluded to in~\cite{Luml70,Luml83}, where it is observed that satisfaction 
of the conventional invariance requirement by the (macroscopic) stress 
is tantamount to ignoring the effects of inertia in the constitutive 
prescription of stress. This observation applies regardless of the
question of invariance of the balance laws themselves.
\section{Invariance: Couples and Angular Momentum Balance}\label{sec:am-inv} 
\par
Returning to the macroscopic angular momentum balance 
equation~\eqref{eq:s-amb1-local}, one may confirm by direct calculation that 
it is intrinsically (that is, without the need for any additional assumptions) 
form-invariant. Furthermore, starting from the respective definitions
in~\eqref{eq:couple-f} and~\eqref{eq:couple-m}, it can be shown with
the aid of~\eqref{eq:g},~\eqref{eq:srbm},~\eqref{eq:srbm_v},
\eqref{eq:m-srbm},~\eqref{eq:m-srbm_v},~\eqref{eq:b+}, as well 
as the invariance of microscopic stress, mass density, and volume, that  
\begin{align}\label{eq:couple-f+}
{\rho^M}^+{\gb^M}^+ & \ =\
\int_{\cR^+} (\xb^+-\yb^+)\times{\rho^m}^+{\bb^m}^+ g^+\,\dm {v^m}^+ \notag \\ 
& \ =\ \Qb\rho^M\gb^M + 
\Qb \int_\Rc (\xb-\yb)\times\rho^m (2\Omegab\vb^m  + \Qb^T\ddot\Qb \xb)g\,\dm v^m
\end{align}
and
\begin{align}\label{eq:couple-m+}
{\Mb^M}^+ & \ =\
\int_{\cR^+} (\xb^+-\yb^+)\times{\Tb^m}^+ g^+\,\dm {v^m}^+ - 
\int_{\cR^+} (\xb^+-\yb^+)\times
\bigl[{\rho^m}^+{\vb^m}^+ \otimes ({\vb^m}^+-{\vb^M}^+)\bigr]
g^+\,\dm {v^m}^+ \notag \\
& \ =\ 
\Qb\Mb^M\Qb^T - \notag \\
   & \hspace{-0.2in} \Qb \int_\Rc (\xb-\yb) \times \rho^m 
\Bigl[(\Omegab\xb + \Qb^T\dot{\cb})\otimes\bigl[(\vb^m-\vb^M) +
\Omegab(\xb-\yb)\bigr] + \vb^m\otimes\Omegab(\xb-\yb) 
\Bigr]g\,\dm v^m\Qb^T\ .
\end{align}
\par
The preceding two equations demonstrate that neither the body couple
nor the couple stress is objective, which is entirely reasonable
given their physical meaning. Again, it is easy to show that 
the couple stress would be objective if the contribution of the
velocity fluctuations can be ignored in~\eqref{eq:couple-m}.
\section{Invariance: Macroscopic Heat Flux and Energy Balance}\label{sec:heat-inv} 
\par
Under superposed rigid-body motions, the macroscopic heat flux vector
in~\eqref{eq:m-heat-flux} is given by 
\begin{equation}\label{eq:macroheatrot}
{\qb^M}^+\ =\ \int_{\cR^+} \left[ {\qb^m}^+ + 
{\Tb^{m^+}}({\vb^m}^+-\mbox{$\hat{\vb}^m$}^+) + 
{\rho^m}^+ \mbox{${\hat{e}^m}$}^+ ({\vb^m}^+ - {\vb^M}^+)
\right]g^+\,\dm {v^m}^+\ . 
\end{equation}
Taking into consideration the invariance properties of the microscopic
stress and heat flux, and invoking~\eqref{eq:g},~\eqref{eq:m-heat-flux},
(\ref{eq:srbm},\ref{eq:srbm_v}), (\ref{eq:m-srbm},\ref{eq:m-srbm_v}), 
and~\eqref{eq:v-vh+}, the preceding expression leads to 
\begin{equation}\label{eq:macroheatrot-1}
{\qb^M}^+\ =\ \Qb\qb^M  +
\Qb\left[\Omegab \int_\cR \rho^m \mbox{${\hat{e}^m}$}^+ (\xb - \yb) g\,\dm v^m +
\int_\cR \rho^m (\mbox{${\hat{e}^m}$}^+ -\hat{e}^m) 
(\vb^m - \vb^M) g\,\dm v^m\right]\ .
\end{equation}
As with the Cauchy stress, it is seen from~\eqref{eq:macroheatrot-1} 
that the macroscopic heat flux is not invariant under superposed rigid-body 
motions, as previously observed~\cite{Muel72,True76,Hoov81}. However, 
unlike stress, the non-objective parts of the heat flux 
in~\eqref{eq:macroheatrot-1} are neither individually nor jointly 
divergence-free relative to the coordinates in the superposed configuration.
\par
As with the balances of mass, linear momentum, and angular momentum, 
the macroscopic balance of energy is form-invariant. This can be
argued in a straightforward manner by invoking the form-invariance 
of the total energy in~\eqref{eq:ext4} and repeating the derivation 
of the energy balance equation contained in Appendix~\ref{sec:app-B} 
using the superposed configuration while exploiting the form-invariance of 
the microscopic energy balance~\eqref{eq:micro_ener}. 
%
%
\section{Conclusions}\label{sec:conclusions} 
\par
The continuum-to-continuum homogenization theory inspired by the
Irving--Kirkwood procedure gives rise to a polar macroscopic medium due
to the length scale inherent in the coarse-graining process. The role
of macroscopic angular momentum becomes non-trivial and a suitable 
definition of the local macroscopic spin enables the additive
decomposition of the total internal energy into non-inertial, 
translational and rotational components, thus enabling a canonical
representation of the contributions of internal forces and stresses 
(both polar and non-polar) in the macroscopic balance of energy. 
\par
The assumption of form-invariance of the extensive relations for mass, 
linear and angular momenta, and total energy combined with the
standard invariance properties in the microscale 
suffices in translating the form-invariance of the microscopic balance
laws to the macroscale, thereby providing a sound theoretical foundation for 
future development of macroscopic models. At the same time, the
homogenization theory yields macroscopic stresses and heat fluxes that
do not observe the conventional invariance requirement due to presence
of inertial effects. These departures, which have been long observed 
in fluctuation-dominated problems, such as turbulent flows, 
are now placed within the realm of a continuum-mechanical theory. 
\par
In broader terms, the paper demonstrates that continuum-to-continuum 
homogenization may be an effective vehicle for investigating (and,
hopefully, expanding) the boundaries of traditional continuum-mechanics, 
as motivated by the study of inhomogeneous materials, through
physically motivated and mathematically prescribed
concepts such as inertial stress and heat flux, body and surface 
couples, and local angular velocity and time-evolving moment-of-inertia 
tensors. Whereas single-scale polar theories may {\em postulate\/}
the existence and evolution of such quantities, the proposed approach
relies on the underlying continuum-mechanical microscale and the
proposed homogenization theory to constitutively specify them.
%
%

$\\${\bf Acknowledgements.} B.E. Abali's work was supported by a grant from the Max 
Kade Foundation to the University of California, Berkeley. K.K. Mandadapu is supported by the University of California, Berkeley.

%
%
\protect\vspace{0.2in}
\par\noindent

\dsp
\newpage
\begin{appendices} 
\numberwithin{equation}{section}

\def\theequation{\Alph{section}.\arabic{equation}}
\section{Derivation of the Macroscopic Energy Balance Equation}\label{sec:app-A}
\par
To derive the macroscopic energy balance equation~\eqref{eq:ma-energy-balance}, 
start by taking the material time derivative of the extensivity
equation~\eqref{eq:ext4} and then invoke~\eqref{eq:micro_mass} and
\eqref{eq:macro_mass} to find that 
\begin{equation}\label{eq:A1} 
\rho^M \dot{e}^M\ =\ 
\rho^M e^M \frac{\partial}{\partial\yb}\cdot\vb^M + 
\int_\Rc \rho^m \dot{e}^m g\, \dm v^m + 
\int_\Rc \rho^m e^m \dot{g}\, \dm v^m\ .
\end{equation}
The second term on the right-hand side of~\eqref{eq:A1} may be
expanded with the aid of the microscopic energy
balance~\eqref{eq:micro_ener} and the definition of macroscopic body
force~\eqref{eq:bodyM} as 
\begin{multline}\label{eq:A1-1} 
\int_\Rc \rho^m \dot{e}^m g\, \dm v^m\ =\ 
\rho^M\bb^M\cdot\vb^M + 
\int_\Rc \rho^m\bb^m\cdot(\vb^m-\vb^M)g\, \dm v^m +
\int_\Rc \rho^m r^m g\, \dm v^m + \\
\int_\Rc \frac{\partial}{\partial \xb}\cdot\left(
\Tb^m\vb^m \right) g\, \dm v^m -  
\int_\Rc \frac{\partial}{\partial \xb}\cdot\qb^m g\, \dm v^m\ .
\end{multline}
However, the divergence theorem, in conjunction with~\eqref{eq:g-boundary}
and~\eqref{eq:iden}, implies that
\begin{equation}\label{eq:A1-1-1}
\int_\Rc \frac{\partial}{\partial \xb}\cdot
\left(\Tb^m\vb^m\right) g\, \dm v^m\ =\ 
\frac{\partial}{\partial \yb}\cdot\int_\Rc \Tb^m(\vb^m-\vb^M)g\, \dm v^m 
+ \frac{\partial}{\partial \yb}\cdot\left[\int_\Rc \Tb^m g\, \dm
v^m \vb^M\right] 
\end{equation}
and
\begin{equation}\label{eq:A1-1-2}
\int_\Rc \frac{\partial}{\partial \xb}\cdot\qb^m g\, \dm v^m\ =\ 
\frac{\partial}{\partial \yb}\cdot\int_\Rc \qb^m g\, \dm v^m\ .
\end{equation}
Likewise, upon using~\eqref{eq:iden} and~\eqref{eq:ext4}, 
the third term on the right-hand side of~\eqref{eq:A1} becomes 
\begin{equation}\label{eq:A1-2} 
\int_\Rc \rho^m e^m \dot{g}\, \dm v^m\ =\ 
-\frac{\partial}{\partial\yb}\cdot
\int_\Rc \rho^m e^m (\vb^m-\vb^M)g\, \dm v^m - 
\rho^M e^M \frac{\partial}{\partial\yb}\cdot\vb^M\ .
\end{equation}
Inserting~\eqref{eq:A1-1} and~\eqref{eq:A1-2} into~\eqref{eq:A1}, and
taking into account~\eqref{eq:A1-1-1},~\eqref{eq:A1-1-2}, and the
definition of the macroscopic Cauchy stress in~\eqref{eq:macrostress}
leads to 
\begin{multline}\label{eq:A2} 
\rho^M \dot{e}^M\ =\ 
\rho^M\bb^M\cdot\vb^M + 
\int_\Rc \bigl[ \rho^m r^m + \rho^m\bb^m\cdot(\vb^m-\vb^M)\bigr]g\, \dm v^m 
+ \frac{\partial}{\partial \yb}\cdot\left(\Tb^M\vb^M\right) \\
-\frac{\partial}{\partial \yb}\cdot\int_\Rc\bigl[ 
\qb^m - \Tb^m(\vb^m-\vb^M) + \rho^m e^m (\vb^m-\vb^M) \bigr]g\, \dm v^m + 
\frac{\partial}{\partial \yb}\cdot\Bigl[\int_\Rc \rho^m
(\vb^m-\vb^M)\otimes(\vb^m-\vb^M) g\, \dm v^m\vb^M\Bigr]\ .  
\end{multline}
\par
To extract the polar effects from the preceding statement of energy
balance, recall the definition of the convected microscopic velocity 
$\hat{\vb}^m$ in~\eqref{eq:v-convected} and note that
\begin{equation}\label{eq:A2-1} 
\int_\Rc \rho^m\bb^m\cdot(\vb^m-\vb^M) g\, \dm v^m\ =\ 
\int_\Rc \rho^m\bb^m\cdot(\vb^m-\hat{\vb}^m) g\, \dm v^m + 
\rho^M\gb^M\cdot\wb^M\ ,
\end{equation}
where use is made of~\eqref{eq:couple-f}. Likewise, it can be shown 
with the aid of~\eqref{eq:couple-m} that
\begin{multline}\label{eq:A2-2} 
\int_\Rc \Tb^m(\vb^m-\vb^M) g\, \dm v^m\ =\ 
\int_\Rc \Tb^m(\vb^m-\hat{\vb}^m) g\, \dm v^m + 
(\Mb^M)^T\wb^M \\ + \left[\int_\Rc (\xb-\yb)\times
\left[\rho^m\vb^m\otimes(\vb^m-\vb^M)\right]g\, \dm v^m\right]^T\wb^M\ ,
\end{multline}
Lastly, upon taking into account the definition of~$\hat{e}^m$ in
\eqref{eq:e-hat}, hence its implied relation to the total internal
energy~$e^m$, the internal energy term in~\eqref{eq:A2} may be expanded into 
\begin{multline}\label{eq:A2-3} 
\int_\Rc \rho^m e^m (\vb^m-\vb^M)g\, \dm v^m\ =\ 
\int_\Rc \rho^m \hat{e}^m (\vb^m-\vb^M)g\, \dm v^m \\ + 
\int_\Rc \frac{1}{2} \rho^m \bigl[\vb^m\cdot\vb^m - 
(\vb^m-\hat{\vb}^m)\cdot(\vb^m-\hat{\vb}^m) -
\hat{\vb}^m\cdot\hat{\vb}^m\bigr](\vb^m-\vb^M)g\, \dm v^m\ .
\end{multline} 
\par
The macroscopic energy balance equation~\eqref{eq:ma-energy-balance}
is obtained by substituting~(\ref{eq:A2-1}-\ref{eq:A2-3}) into
\eqref{eq:A2} and using~\eqref{eq:v-convected} to eliminate all
residual terms.
\section{Divergence-free Terms in the Macroscopic Cauchy Stress}\label{sec:app-B}
\par
Preliminary to establishing the divergence-free property of the additional 
inertial terms in~\eqref{eq:macrostress_srbm_3}, two useful identities 
are deduced. For the first identity, start by taking the material time 
derivative of the invariance relation~\eqref{eq:g}, which yields 
\begin{equation} \label{eq:grel-1}
\gradyg \cdot \vb^{M} + \gradxg \cdot \vb^{m}\ =\ 
\gradypg \cdot {\vb^M}^+ + \gradxpg \cdot {\vb^m}^+\ .
\end{equation}
Next, upon invoking~\eqref{eq:srbm},~\eqref{eq:srbm_v},~\eqref{eq:m-srbm}
and~\eqref{eq:m-srbm_v}, equation~\eqref{eq:grel-1} may 
be rewritten as 
\begin{equation}
\gradyg \cdot \vb^{M} + \gradxg \cdot \vb^{m} \\
=\ \gradyg \cdot (\vb^M + \Omegab\yb) + \gradxg \cdot (\vb^m +
\Omegab\xb) + \Qb\left(\gradyg + \gradxg\right)\cdot\dot{\cb} 
\end{equation}
and further reduced, upon observing~\eqref{eq:iden}, to 
\begin{equation}\label{eq:grel_a}
\Omegab \cdot \Big( \xb \otimes \gradxg + \yb \otimes \gradyg \Big)\ =\ 0\ .
\end{equation}
Given the arbitrariness of~$\Omegab$, the preceding
equation implies that the quantity in parentheses is necessarily symmetric. 
Furthermore, upon using again~\eqref{eq:iden}, equation~\eqref{eq:grel_a} 
readily implies the first identity, in the form  
\begin{equation}\label{eq:grel_a1}
\Omegab(\xb - \yb)\cdot\gradyg\ =\ 0\ .
\end{equation}
\par
The second identity is obtained by taking the material time derivative 
of the center-of-mass relation~\eqref{eq:cenmass-1}.
To this end, appealing to the Reynolds transport theorem and using 
the microscopic balance of mass~\eqref{eq:micro_mass}, it follows that 
\begin{equation}\label{eq:grel-2}
\frac{\dm}{\dm t} ({\rho^M} \yb)\ =\ 
\int_{\cR} \bigg[{\rho^m} {\vb^m} g + {\rho^m} \xb 
\left(\gradxg \cdot {\vb^m}\right) + 
{\rho^m} \xb \left(\gradyg \cdot {\vb^M}\right) \bigg]\,\dm {v^m}\ . 
\end{equation}
Using first~\eqref{eq:iden} and then invoking~\eqref{eq:ext2} 
and~\eqref{eq:cenmass-1}, the preceding equation becomes 
\begin{align}\label{eq:grel-2a} 
\frac{\dm}{\dm t} (\rho^M \yb)\ & =\ 
\int_{\cR} {\rho^m} {\vb^m} g \,\dm {v^m} - 
\grady \cdot \int_{\cR} {\rho^m} \xb \otimes
({\vb^m} - {\vb^M}) g\,\dm {v^m} - \int_{\cR}{\rho^m} \xb\, \gradyg \cdot {\vb^M}\,\dm {v^m} 
\notag \\
& =\ {\rho^M} {\vb^M} - \grady \cdot \int_{\cR}
{\rho^m} (\xb -\yb) \otimes ({\vb^m}-{\vb^M})g\, \dm {v^m} 
-{\rho^M}\yb\, \grady \cdot {\vb^M}  \ . 
\end{align}
Expanding now the left-hand side of~\eqref{eq:grel-2}, 
and using the macroscopic mass balance equation~\eqref{eq:macro_mass},
it is concluded from~\eqref{eq:grel-2a} that 
\begin{equation}\label{eq:grel_b}
\grady \cdot \int_{\cR} {\rho^m} (\xb -\yb) \otimes
({\vb^m}-{\vb^M}) g \, \dm {v^m} \ = \ \zerob \ ,
\end{equation}
which is the second identity of interest here. 
\par
It is now possible to show that the last three terms on the right-hand side of 
\eqref{eq:macrostress_srbm_3} are individually divergence-free.  Indeed, 
consider the first term, which takes the form 
\begin{equation}\label{eq:divA-1}
\begin{split}
\gradyp \cdot (\Qb \Omegab \Ab \Qb^T) & \ =\ \Qb \grady \cdot (\Omegab \Ab) \\
 & \ =\ \Qb \Omegab  
\grady \cdot \int_{\cR} {\rho^m} (\xb -\yb) \otimes (\vb^m-\vb^M) g\,\dm {v^m}^+
\end{split}
\end{equation}
and vanishes identically due to~\eqref{eq:grel_b}. The next term is 
\begin{equation}\label{eq:divA-2}
\begin{split}
\grady \cdot \bigl(\Qb (\Omegab \Ab)^T \Qb \bigr) & \ =\ 
\Qb \grady \cdot (\Omegab \Ab)^T \\
  & \ =\  \Qb \grady \cdot \int_\cR \rho^m ({\vb^m}-{\vb^M}) \otimes 
 (\xb -\yb) g\,\dm v^m \Omegab^T  \\ 
& \ =\ 
\Qb\int_{\cR} {\rho^m} \left(-\frac{\partial{\vb^M}}{\partial\yb}\Omegab\right) 
(\xb -\yb)g \dm {v^m} \\
& \hspace{1.0in} + 
\Qb \int_{\cR} {\rho^m} ({\vb^m}-{\vb^M}) 
(-\ib\cdot\Omegab)g \dm {v^m}  \\
& \hspace{1.25in} 
+ \Qb \int_{\cR} {\rho^m} ({\vb^m}-{\vb^M}) 
\bigl[\Omegab(\xb -\yb)\cdot\gradyg\bigl]\, \dm {v^m}
\end{split}
\end{equation}
The three terms on the right-hand side of~\eqref{eq:divA-2} themselves vanish 
individually due to \eqref{eq:cenmass-1}, the skew-symmetry of~$\Omegab$, and
the identity~\eqref{eq:grel_a1}, respectively. 
Lastly, given the definition of~$\Bb$ in~\eqref{eq:B}, one may write 
\begin{equation}\label{eq:divB-1}
\begin{split}
\grady \cdot (\Qb\Omegab \Bb \Omegab^T\Qb^T) & \ =\ 
\Qb \grady\cdot(\Omegab \Bb \Omegab^T) \\
& \ =\ \Qb \grady \cdot \int_{\cR} {\rho^m} \Omegab (\xb-\yb) 
\otimes \Omegab (\xb -\yb) g\, \dm {v^m} \\ 
&\ =\ 
-\Qb \int_{\cR} {\rho^m} \Omegab^2 (\xb -\yb)g \dm {v^m} \\
&\hspace{1.0in} - 
\Qb \int_{\cR} {\rho^m} \Omegab (\xb-\yb) 
(-\ib\cdot\Omegab)g \dm {v^m}  \\
& \hspace{1.25in} 
+ \Qb \int_\cR \rho^m \Omegab(\xb-\yb) 
\bigl[\Omegab(\xb -\yb)\cdot \gradyg\bigr]\, \dm v^m\ .
\end{split}
\end{equation}
Again, each of the three terms on the right-hand side of~\eqref{eq:divB-1} 
vanishes owing to~\eqref{eq:cenmass-1}, the skew-symmetry of
$\Omegab$, and the identity~\eqref{eq:grel_b}, respectively. 
Therefore, the last three terms on the right-hand side of
\eqref{eq:macrostress_srbm_3} are individually divergence-free with
respect to the macroscopic coordinates in the rigidly transformed frame. 
\end{appendices} 
\newpage
%
%
\clearpage

\end{document}